\def\refbib{\hangindent\parindent\noindent}
\def\ts{\thinspace}

\def\1{\'\i }

\newcommand{\gtsim}{\lower.5ex\hbox{$\; \buildrel > \over \sim \;$}}
\newcommand{\ltsim}{\lower.5ex\hbox{$\; \buildrel < \over \sim \;$}}

\def\Z{\sf Z}
\def\S{Sco{\ts}X-1}
\def\9{GX{\ts}349$+$2}
\def\0{GX{\ts}340$+$0}
\def\5{GX{\ts}5$-$1}
\def\7{GX{\ts}17$+$2}
\def\C{Cyg{\ts}X-2}
\def\M{$\dot{M}$}
\def\flu#1#2{#1{\ts}$\times${\ts}$10^{-#2}${\ts}photons{\ts}cm$^{-2}${\ts}s$^{-1}$}

\documentclass[cmfonts]{aipproc}
\usepackage{times}

\layoutstyle{6x9}

\usepackage[dvips]{epsfig}
\usepackage{graphicx}

\begin{document}

\title[Hard X--ray Tails in {\sf Z} Sources]
      {Studies of Hard X--ray Tails in {\sf Z} Sources with HEXTE/RXTE}

\author{Flavio D'Amico}
{
       address={Center for Astrophysics and Space Sciences, University of California, San Diego, and \\
                Instituto Nacional de Pesquisas Espaciais - INPE \\
                Av. dos Astronautas 1758, 12227-010 S. J. dos Campos, Brazil},
       email={fdamico@mamacass.ucsd.edu},
}

\author{William A. Heindl}
{ 
       address={Center for Astrophysics and Space Sciences, University of California, San Diego \\
                9500 Gilman Dr., La Jolla, CA 92093-0424},
       email={wheindl@ucsd.edu},
}  

\author{Richard E. Rothschild}
{ 
       address={Center for Astrophysics and Space Sciences, University of California, San Diego \\
                9500 Gilman Dr., La Jolla, CA 92093-0424},
       email={rrothschild@ucsd.edu},
}  

\author{Duane E. Gruber}
{ 
       address={Center for Astrophysics and Space Sciences, University of California, San Diego \\
                9500 Gilman Dr., La Jolla, CA 92093-0424},
       email={dgruber@ucsd.edu}
}               

\begin{abstract}
We report {\it RXTE} results of spectral analyses of three   ({\S}, {\9}, 
and {\C}) out of the 6 known {\Z} sources. No hard X--ray tails were found
for {\C} ($< \,$\flu{8.4}{5}, 50--100{\ts}keV, 3$\sigma$)
and for {\9} ($< \,$\flu{7.9}{5}, 50--100{\ts}keV, 3$\sigma$). For {\S} a variable 
hard X--ray tail (with an average flux of \flu{2.0}{3},
50--100{\ts}keV) has already been reported. We compare our results to 
reported detections of a hard component in the spectrum of {\C} and {\9}. We
argue that, taking into account all the results on  detections of hard X--ray
tails in {\S} and {\9}, the appearance of such a component is correlated 
with the brightness of the thermal component.  

\end{abstract}

\maketitle

\section{Introduction}

The class of {\Z} sources comprises 6 LMXBs ({\S}, {\9}, {\0}, {\C}, 
{\5}, and {\7}) in which the primary is a neutron star with a low magnetic 
field ($\sim${\ts}$10^{10}${\ts}G) accreting at or near the Eddington
limit [1]. They share similar timing properties and are among the most 
luminous known LMXBs. The designation {\Z} source  results from the shape 
described  in a x--ray color{\ts}$\times${\ts}color diagram (CD), with the 
movement along the {\Z} interpreted in terms of changes in the mass accretion 
rate ({\M}, see, e.g., [1]). Apart from {\S} and {\C}, the {\Z} sources are 
all found near the Galactic mid-plane (i.e., $b \, = \, 0^{\circ}$).

Hard X--ray spectra from both {\Z} and low luminosity atoll sources have already 
been reported in the literature [2-8]. The production of the hard X--ray tails 
in atoll sources has been presented in the context of various thermal emission 
models [9] from which the accretion geometry can be inferred. The situation is 
less clear for the {\Z} sources, where non--thermal mechanisms are invoked to 
explain the production of such a component, and little, or nothing, is known 
about the details of the accretion geometry.

We are currently analyzing all of the {\Z} source observations 
in the public {\it RXTE} database which contain long pointings. 
The aim is to create an uniform database that will allow
us to make direct hard X--ray spectra comparisons. From this
we expect to better understand the behavior of  any non--thermal 
emission in these sources. We report here the preliminary results of 
this work, with data from 3 ({\S}, {\9}, and {\C}) out of the 6 
known {\Z} sources.

\section{Data Selection and Analysis}

We used data from HEXTE [10] to search for hard X--ray tails in the spectrum 
of {\S}, {\9}, and {\C} in the  $\sim${\ts}20--220{\ts}keV interval and data 
from PCA [11] to determine the position of the source in the CD and to study 
the 2--20{\ts}keV spectrum. We selected, from the public {\it RXTE} database, 
those subsets of data in which $\gtsim${\ts}5000{\ts}s of HEXTE total 
on--source time was available, in order to achieve good sensitivity at high 
energies. Table{\ts}{\ref{tab1}} shows the selected subsets for {\9} and {\C}. 
The list of selected observations of {\S} is given in [7]. We  used XSPEC
to analyze the PCA source spectra, using published models for {\9}
(a blackbody plus a disk-blackbody and an iron line, see [12]) and
{\C} (an absorbed cutoff power-law plus an iron line, see [13]). A
complex multicomponent model (an absorbed blackbody plus a power-law, a
Comptonization spectrum, and a Gaussian line) was used to heuristically fit 
the PCA {\S} spectra. Low enewrgy (20--50{\ts}keV) HEXTE spectra were fitted 
by a simple thermal bremsstrahlung. The hard X-ray component 
(i.e. $E > 50${\ts}keV), found only in {\S}, was modeled as a simple
power-law (see [6] for a more detailed description of the instrument and 
procedure used for data analysis). We carefully verified our background 
subtraction procedures, specially for {\9}, which is located near the Galactic 
mid-plane, where the diffuse Galactic Plane background up
to $E \sim$ 800{\ts}keV [14] is known to vary in latitude [15].  We took 
advantage of HEXTE aperture modulation to remove this contribution to the 
background since HEXTE Cluster A measured the background at the same latitude 
as the source.  Source confusion is also a concern for {\9} due to the presence 
of 4U{\ts}1700$-$37 (see, e.g., [16]) inside the field of view of one of the 
regions used by HEXTE Cluster B to measure background (the B$^-$ region). This 
is easily solved using only the B$^+$ region to measure the background for HEXTE
Cluster B.  We found no evidence of source confusion/contamination for {\C} and {\S}.

\begin{table}
\begin{tabular}{cccccccc}
\hline
{\hspace{-1truecm}{\bf GX 349+2}}
               &        &       &      &      &      &      &                                        \\
{\bf OBSID}                                                                              & 
{\bf MJD}                                                                                & 
{\bf T$_{obs}$}\tablenote{total {\it RXTE} source's exposure time, in s}                 & 
{\bf T$_{HEX}$}\tablenote{corrected HEXTE exposure time, in s}                           & 
{\bf F$_{(2-20)}$}\tablenote{Flux, in 2-20 keV range, in units of
10$^{-8}$ ergs cm$^{-2}$ s$^{-1}$; uncertainties are given at 90{\%} confidence level}   & 
{\bf F$_{(20-50)}$}\tablenote{Flux, in 20-50 keV range, in units of
10$^{-10}$ ergs cm$^{-2}$  s$^{-1}$, for GX{\ts}349$+$2, and 
10$^{-11}$ ergs cm$^{-2}$  s$^{-1}$, for Cyg{\ts}X-2; 
uncertainties are given at 90{\%} confidence level}                                      & 
{\bf F$_{(50-100)}$}\tablenote{3$\sigma$ upper limit on power-law Flux, in units of
10$^{-11}$ ergs cm$^{-2}$  s$^{-1}$, in the 50-100 keV range; power-law index
frozen at a value of 2}                                                                  &
{\bf {\Z}}\tablenote{HB=horizontal branch; NB=normal branch; FB=flaring
branch; SA=soft apex}                                                                                 \\
20054-05-01-00 & 50570 & 10032 & 5902 & 1.75$^{+0.09}_{-0.09}$ & 2.03$^{+0.43}_{-0.41}$  & $< 4.64$
                                                                                         & SA         \\
30042-02-01-01 & 50822 &  8688 & 5492 & 2.42$^{+0.22}_{-0.17}$ & 3.94$^{+0.59}_{-0.52}$  & $< 5.52$
                                                                                         & FB         \\
30042-02-01-02 & 50823 & 10336 & 6527 & 1.98$^{+0.06}_{-0.06}$ & 3.46$^{+0.48}_{-0.48}$  & $< 6.28$
                                                                                         & (lower) NB \\
30042-02-01-07 & 50823 & 14160 & 8850 & 1.95$^{+0.02}_{-0.02}$ & 2.87$^{+0.34}_{-0.34}$  & $< 1.37$
              										 & SA         \\
30042-02-01-03 & 50825 & 13728 & 8602 & 2.50$^{+0.25}_{-0.22}$ & 3.94$^{+0.08}_{-0.83}$  & $< 4.84$
											 & FB         \\
30042-02-01-04 & 50826 & 14304 & 8865 & 2.55$^{+0.20}_{-0.18}$ & 2.56$^{+0.28}_{-0.26}$  & $< 3.71$
											 & FB         \\
30042-02-01-08 & 50826 & 10368 & 6318 & 2.08$^{+0.25}_{-0.25}$ & 4.97$^{+0.45}_{-0.45}$  & $< 3.17$
 											 & FB         \\
30042-02-02-00 & 50830 &  9760 & 5632 & 1.72$^{+0.07}_{-0.07}$ & 2.46$^{+0.44}_{-0.44}$  & $< 5.32$
											 & NB-FB      \\
30042-02-02-08 & 50838 &  7704 & 4689 & 1.71$^{+0.08}_{-0.08}$ & 2.10$^{+0.48}_{-0.46}$  & $< 6.49$
											 & NB-FB      \\
30042-02-03-01 & 50842 &  9216 & 5684 & 2.71$^{+0.49}_{-0.32}$ & 4.22$^{+0.51}_{-0.46}$  & $< 2.97$
											 & FB         \\
\hline
{\hspace{-1.3truecm}{\bf Cyg X-2}}  
               &       &       &      &                        &                         &     &      \\
10063-10-01-00 & 50316 &  8088 & 5044 & 1.16$^{+0.35}_{-0.39}$ & ~8.37$^{+4.77}_{-4.44}$ & $< 6.92$ 
                                                                                         & FB         \\   
30418-01-05-00 & 51000 & 10760 & 6349 & 1.14$^{+0.14}_{-0.18}$ & ~5.62$^{+4.22}_{-3.82}$ & $< 4.12$
											 & FB         \\
30046-01-01-00 & 51009 & 13376 & 8180 & 1.88$^{+0.13}_{-0.13}$ & 21.89$^{+3.50}_{-3.28}$ & $< 4.61$
											 & SA         \\
30046-01-02-00 & 51015 & 14736 & 9240 & 1.65$^{+0.07}_{-0.15}$ & 15.26$^{+3.66}_{-3.66}$ & $< 4.21$
											 & FB         \\
30046-01-03-00 & 51022 & 13728 & 8881 & 1.42$^{+0.30}_{-0.28}$ & 28.03$^{+3.64}_{-3.64}$ & $< 4.88$
											 & FB         \\
30046-01-04-00 & 51029 & 13584 & 8157 & 1.23$^{+0.09}_{-0.10}$ & ~9.74$^{+3.12}_{-3.02}$ & $< 1.78$
											 & FB         \\
30046-01-06-00 & 51041 & 15104 & 9114 & 1.76$^{+0.18}_{-0.21}$ & 15.56$^{+3.11}_{-3.11}$ & $< 4.86$
											 & FB         \\
30046-01-07-00 & 51048 & 13888 & 8231 & 1.35$^{+0.09}_{-0.11}$ & ~5.38$^{+3.34}_{-3.12}$ & $< 3.02$
											 & FB         \\
30046-01-08-00 & 51055 & 13920 & 8566 & 1.20$^{+0.06}_{-0.06}$ & 25.95$^{+3.37}_{-3.37}$ & $< 4.40$
											 & NB         \\
30046-01-09-00 & 51061 & 16256 & 9010 & 1.30$^{+0.08}_{-0.08}$ & ~6.54$^{+3.07}_{-2.88}$ & $< 3.62$
											 & FB         \\
30046-01-10-00 & 51068 &  8600 & 5419 & 1.81$^{+0.11}_{-0.13}$ & 23.97$^{+5.03}_{-5.03}$ & $< 3.03$
											 & SA         \\
30046-01-11-00 & 51078 & 12512 & 8098 & 1.48$^{+0.10}_{-0.12}$ & ~9.24$^{+3.88}_{-3.60}$ & $< 3.96$
											 & FB         \\
30046-01-12-00 & 51081 & 14608 & 9703 & 1.37$^{+0.01}_{-0.01}$ & 29.16$^{+3.50}_{-3.50}$ & $< 7.01$
											 & HB         \\
\hline
\end{tabular}
\caption{Selected {\it RXTE} observations of GX{\ts}349$+$2 and Cyg{\ts}X-2}
\label{tab1}
\end{table}

\section{Results}

\begin{figure}
\epsfig{file=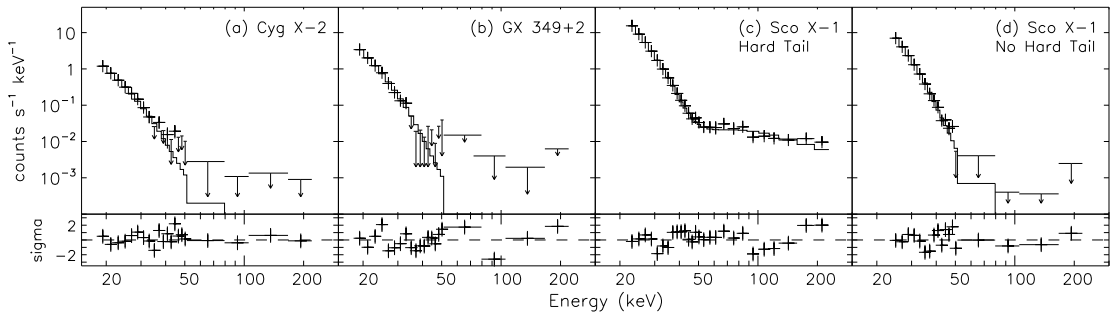, width=4.3cm, angle=90}
\caption{Typical HEXTE spectra (upper panels) for (a) {\C}, (b) {\9}, 
         (c) a hard X-ray tail detection in {\S}, and (d) a non-detection
         in {\S} (for comparison). Residuals are given in units of standard 
         deviations (lower panels). Upper limits are 2$\sigma$.}
\label{fig1}
\end{figure}

Cygnus{\ts}X-2 and {\9} were easily detected by HEXTE up to
50{\ts}keV.  Nevertheless, the detection level was {\it always}
below 3$\sigma$ in the 50--75{\ts}keV band.
We show in Fig. {\ref{fig1}} a typical spectrum for {\C} and {\9} together
with a detection and a non-detection of a hard X--ray tail in {\S}.

All sources show some degree of variability in the 20-50{\ts}keV range. 
From the results in [7], for  {\S}, a factor of 2 was detected, while it 
was a factor of 5  for {\C} and 2 for {\9}. Among the three, {\C} is the 
least luminous in the 2--20{\ts}keV  energy range, with an average luminosity 
of 0.4{\ts}$L_{\hbox{\footnotesize{Edd}}}$ (using $d$ and $M_{\hbox{ns}}$
measurements in [17]), while {\S} and {\9} emit at or above Eddington levels, for  
$M_{\hbox{ns}} = 1.4 \,  M_{\odot}$ (see [18] and [19] for measured distances to 
{\S} and {\9}, respectively). We found no evidence of the presence of a hard X--ray 
tail in our database for {\9} or {\C}. The HEXTE 3${\sigma}$ upper limit to 
50--100{\ts}keV flux from {\9} is \flu{7.9}{5} and for {\C} is \flu{8.4}{5}.
For these two sources, a hard X--ray tail was, however, reported by 
{\it BeppoSAX} ([8] and [3], respectively), at a level of \flu{4.6}{4} for {\9} 
(using the fit parameters given in [8]; for {\C} it is not possible to estimate the 
flux from [3]). Our results, thus, can be interpreted in terms of variability
in the appearance of this component, as was observed in {\S} [7] 
on a 4 hour time-scale. 

\section{Discussion}

Scorpius{\ts}X-1 remains as a special case among the {\Z}
sources. It is the only one in which a hard X--ray tail has been
observed more than once, and by two different instruments ([5] and
[7]). For {\C}, {\7} and {\9} hard X--ray  tails were reported by {\it
BeppoSAX} ([3], [4], and [8], respectively) on one occasion. From our
combined HEXTE database, we found the presence of a hard X--ray tail
in 8 out of 28 occasions for {\S}, and zero out of 10 and 13
observations of {\9} and {\C}, respectively. Fitting our HEXTE data
for {\9} and {\C} with a power-law with indices frozen in the range
1--2 (within the values found for those three sources: see [3],
[7-8]), we found a 3$\sigma$ upper limit on the luminosity of the power-law
component, $L^{\hbox{\tiny{PL}}}_{\hbox{\tiny{20-80
keV}}} = 6.8 \, \times \, 10^{35}${\ts}ergs{\ts}s$^{-1}$ and  
$L^{\hbox{\tiny{PL}}}_{\hbox{\tiny{20-80
keV}}} = 5.0 \, \times \, 10^{35}${\ts}ergs{\ts}s$^{-1}$ for {\9} and {\C},
respectively. Our HEXTE result (for $\Gamma = 1-2$) for hard X--ray tail 
detections in {\S} is $L^{\hbox{\tiny{PL}}}_{\hbox{\tiny{20-80
keV}}} = 6.7 \, \times \, 10^{35}${\ts}ergs{\ts}s$^{-1}$. It thus
appears that our observations were sensitive enough to detect hard X--ray tails
in {\C} and {\9}. As we pointed out in [7] the chance of observing a hard X--ray 
tail (in {\S}) is higher when the thermal component of the
spectrum is brighter. From our results here (see Table {\ref{tab1}}),
we have, for {\9} $L^{\hbox{\tiny{Thermal}}}_{\hbox{\tiny{20-50
keV}}} \,  = \,$ 1.2--3.1{\ts}$\times \, 10^{36}$ergs{\ts}s$^{-1}$, while for {\C} 
the results are 
$L^{\hbox{\tiny{Thermal}}}_{\hbox{\tiny{20-50
keV}}} \, = \,$ 0.4--2.1{\ts}$\times \, 10^{36}$ergs{\ts}s$^{-1}$. The same
component in {\S}, when a hard tail is detected [7], is in the range 
$L^{\hbox{\tiny{Thermal}}}_{\hbox{\tiny{20-50 keV}}}$ \, = \, 4.5--9.0{\ts}$\times \,
10^{36}$ergs{\ts}s$^{-1}$. While comparable values were not given by the
{\it BeppoSAX} results in [3], [4], and [8] (nor by the OSSE/{\it
CGRO} results in [5]), it is possible to extrapolate the results presented
in [8] in order to find an estimate of the luminosity of the thermal 
component. We estimate that the 20--50{\ts}keV {\9} luminosity measured 
by {\it BeppoSAX} was greater than $5 \, \times \,
10^{36}${\ts}ergs{\ts}s$^{-1}$. Thus, one can speculate that the
production of a hard X--ray tail in a {\Z} source is a process triggered when 
the thermal component is brighter than a level of $\sim${\ts}$4 \, \times \,
10^{36}${\ts}ergs{\ts}s$^{-1}$.

\section{Conclusions}

We have shown {\it RXTE} results of broad-band spectral analyses of three
{\Z} sources, with emphasis on the hard X--ray spectrum. We found no
evidence for a detection of a hard X--ray tail in the spectra of {\9}
and {\C}, although  one detection of such a component has been
reported for each of these sources. We interpret this in terms of
variability, which was shown
to be as fast as 4 hours in {\S}. We found an
indication that the production of hard X--ray tails in {\Z} sources is
a process triggered when the thermal component brightness is above a
value of $\sim${\ts}$4 \, \times \, 10^{36}${\ts}ergs{\ts}s$^{-1}$. We are
currently creating a uniform HEXTE database including the other
three {\Z} sources ({\7}, {\0}, and {\5}),
from which we hope to be able to better understand the production of hard X--ray 
in {\Z} sources. 

\section{Acknowledgments}

This research has made use of data obtained through the HEASARC,
provided by NASA/GSFC. F.D. gratefully acknowledges FAPESP/Brazil for 
financial support under grant 99/02352-2. This research was supported 
by NASA contract NAS5-30720.

\section{References}

\refbib 1. van der Klis, M., {\it X-Ray Binaries}, edited by
W. H. G. Lewin, J. van Paradjis, and E. P. J. van den Heuvel, Cambridge
University Press, Cambridge, 1995, pp.252--307.

\refbib 2. Barret, D. et al., {\it ApJ} {\bf 533}, 329-351 (2000).

\refbib 3. Frontera, F. et al., {\it Nucl. Phys. B} \ {\bf 69}, 286-293 (1998).

\refbib 4. Di Salvo, T. et al., {\it ApJ} {\bf 544}, L119-L122 (2000).

\refbib 5. Strickman, M., and Barret, D.  2000, ``Detections of
Multiple Hard X-ray Flares from Sco X-1 with OSSE'', in {\it
Proceedings of the Fifth Compton Symposium}, edited by M. L. McConnel and
J. M. Ryan, AIP Conference Proceedings 510, New York, 2000, pp. 222-226.

\refbib 6. D'Amico, F. et al., {\it ApJ} {\bf 547}, L147-L150 (2001).

\refbib 7. D'Amico, F. et al., {\it Adv. Spa. Res.}, in press (2001)
(astro-ph/0101396).

\refbib 8. Di Salvo, T. et al., {\it ApJ}, in press (2001) (astro-ph/0102299). 

\refbib 9. Barret, D., {\it Adv. Spa. Res.}, in press (2001) (astro-ph/0101295).

\refbib 10. Rothschild, R. E. et al., {\it ApJ} {\bf 496}, 538-549 (1998).

\refbib 11. Jahoda, K. et al., {\it Proc. SPIE} {\bf 2808}, 59-70 (1996).

\refbib 12. Christian, D. J., and Swank, J. H., {\it ApJS} {\bf 109},
177-224 (1997).

\refbib 13. Kuulkers, E. et al., {\it A}{\&}{\it A} {\bf 323}, L29-L32 (1997).

\refbib 14. Boggs, S. E. et al., {\it ApJ} {\bf 544}, 320-329 (2000).

\refbib 15. Valinia, A., and Marshall, F. M., {\it ApJ} {\bf 505},
134-147 (1998).

\refbib 16. Reynolds, A. P. et al., {\it A}{\&}{\it A} {\bf 349},
873-876 (1999).

\refbib 17. Orosz, J., and Kuulkers, E., {\it MNRAS} {\bf 305},
132-142 (1999).

\refbib 18. Bradshaw, C. F., Fomalont, E. B., and  Geldzahler, B. J.,
{\it ApJ} {\bf 512}, L11-L14 (1999).

\refbib 19. McNamara, D. H. et al., {\it Pub. Astr. Soc. Pac.} {\bf
112}, 202-216 (2000).

\end{document}